\documentclass[useAMS,usenatbib]{mn2e}

\usepackage[T1]{fontenc}
\usepackage[latin9]{inputenc}
\usepackage{color}
\usepackage{float}
\usepackage{url}
\usepackage{amsmath}
\usepackage{amssymb}
\usepackage{graphicx}
\usepackage{setspace}
\usepackage{esint}

\title[Sparsely Sampling the Sky: Regular vs Random Sampling]{Sparsely Sampling the Sky: Regular vs Random Sampling}
\author[P. Paykari et. al.]{P. Paykari$^{1}$\thanks{E-mail:
paniez.paykari@cea.fr}
S. Pires$^{1}$
J. -L. Starck$^{1}$
and A. H. Jaffe$^{2}$\\
$^{1}$Laboratoire AIM, UMR CEA-CNRS-Paris 7, Irfu, SAp/SEDI, Service d'Astrophysique, CEA Saclay, F-91191 GIF-
SUR-YVETTE CEDEX, France.\\
$^{2}$Department of Physics, Blackett Laboratory, Imperial College, London SW7 2AZ, United Kingdom}

\begin{document}
\date{Accepted 2012}
\pagerange{\pageref{firstpage}--\pageref{lastpage}} \pubyear{2013}
\maketitle
\label{firstpage}

\begin{abstract}
The next generation of galaxy surveys, aiming to observe millions of galaxies, are expensive both
in time and cost. This raises questions regarding the optimal investment
of this time and money for future surveys. In a previous work, it was shown that a sparse
sampling strategy could be a powerful substitute for the contiguous observations. 
However, in this previous paper a {\it regular} sparse sampling was investigated, where the sparse observed patches were regularly distributed on the sky. The regularity of the mask introduces a periodic pattern in the window function, which induces periodic correlations at specific scales. 
In this paper, we use the Bayesian experimental design to investigate a {\it random} sparse sampling, where the observed patches are randomly distributed over the total sparsely sampled area. 
We find that, as there is no preferred scale in the window function, the induced correlation is evenly distributed amongst all scales. This could be desirable if we are interested in specific scales in the galaxy power spectrum, such as the Baryonic Acoustic Oscillation (BAO) scales. However, for constraining the overall galaxy power spectrum and the cosmological parameters, there is no preference over regular or random sampling. Hence any approach that is practically more suitable can be chosen and we can relax the regular-grid condition for the distribution of the observed patches. 
\end{abstract}


\section{Introduction}

The accurate measurement of the cosmological parameters
relies on accurate measurements of a power spectrum, which describe
the spatial distribution of an isotropic random field. The power spectrum is enough to define the perturbations completely
when the perturbations are assumed uncorrelated Gaussian random fields
in the Fourier space. Power spectra (or its Fourier transform, the correlation function) are
what the surveys actually measure, from which cosmological parameters are inferred. 
These spectra are normally a convolution
of the primordial power spectrum\footnote{The primordial power spectrum measures the statistical distribution
of perturbations in the early universe, for example, just after the inflationary era.} and a transfer function, which
depends on the cosmological parameters. One of the most important observed spatial power spectra is the galaxy power spectrum, which was first formulated by \citet{Peebles1973}. This is defined as 
\begin{equation}
P_{g}(k)=2\pi^{2}\cdot b^{2}(k)\cdot k\cdot T^{2}(k)\cdot P_{p}(k)\;,
\end{equation}
where $P_{p}(k)=A_{s}k^{n_{s}-1}$ is the primordial power spectrum and $T(k)$ is the transfer function, which depends upon the cosmological
parameters (e.g., the matter density $\Omega_m$, the scalar spectral index, $n_s$, etc.) responsible for the evolution of the
universe. The bias $b$ relates the galaxy power spectrum
to the underlying matter power spectrum. The galaxy power spectrum is very rich in
terms of constraining a large range of cosmological parameters. On
large scales it probes the structures which are less affected
by clustering and evolution and hence have a `memory' of the initial state of the universe. On intermediate scales
the spectrum informs us about the evolution of the
universe; for example the epoch of matter-radiation equality. 
On relatively small scales there is a great deal of information about
galaxy clustering via the Baryonic Acoustic Oscillations (BAO), which
encodes information about the sound horizon at the time of recombination.
Therefore, measuring the galaxy power spectrum on a large range of
scales helps us to constrain a range of cosmological parameters.

For accurate measurements of the galaxy power spectrum, surveys aim to maximize the observed number of galaxies to overcome the Poisson noise. 
Considering the large investments
in time and money for these surveys, one would like to know the optimal survey strategy. For example, to investigate larger
scales, it may be more efficient to observe a larger, but sparsely
sampled, area of sky instead of a smaller contiguous area. In this
case one gathers a larger density of states in Fourier space,
but at the expense of an increased correlation between different scales
--- aliasing. This would smooth out features on certain scales and decrease
their statistical significance. 
The sparse sampling approach was investigated in a previous paper \citep{PaykJaffeSSS}, where the advantages and disadvantages of such a design was studied. It was shown that a sparse sampling could be a powerful substitute for a contiguous sampling. In particular, it was shown that for a survey similar to the Dark Energy Survey (DES)\footnote{\url{http://www.darkenergysurvey.org/}}, a sparse design could help reduce the observing time (and hence the cost of the survey),
for the same amount of constraining power for the cosmological parameters. Alternatively, for the same amount of observing time, one can observe a larger, but sparsely sampled, area of the sky to improve the constraining power of the survey. However, in their sparse design, the observed patches were regularly distributed over the total sampled area of the sky. The fixed and determined positions of the patches introduces a periodic pattern in the window function, which induces a periodic aliasing of scales. This causes certain scales, which correspond to the fixed distances between the patches, to be more aliased than others. This regular design may, therefore, not be desirable for two reasons; 1. if we are interested in certain scales, such as the Baryonic Acoustic Scales (BAO) or the scale of matter-radiation equality in the galaxy power spectrum, the regular design may not be preferred due to its periodic induced aliasing; 2. a rigid regular distribution of the observed patches may not be practically feasible, as there are regions we would like to avoid, such as the plane of the Milky Way. 

If one is interested in all scales equally and would like to measure all the scales with the same statistical significance, a random sparse sampling may be the preferred approach. In this case, the patches are randomly distributed over the total sampled area, which is practically more feasible. In this work, we investigate both regular and random sparse sampling.

As in \cite{PaykJaffeSSS} we will make use of the Bayesian
Experimental Design and the Figure of Merit (FoM) to select the optimal design for constraining the galaxy power spectrum bins and a set of cosmological parameters.

\section{Bayesian Experimental Design, Figure-of-Merit and Fisher Matrix Analysis for Galaxy Surveys}

Bayesian methods have recently been used in cosmology for model comparison
and for deriving posterior probability distributions for parameters
of different models. Bayesian statistics can also be used to investigate the performance of future experiments,
based on our current knowledge \citep{Liddleetal06,Trotta07,Trotta07-BayesFactor}.
We will use this strength of Bayesian statistics to optimise the strategy to observe the sky for galaxy
surveys. For such an optimisation, we need to satisfy three requirements: 
1. specify the parameters that define the experiment; 
2. specify the parameters to constrain (with respect to which the survey is optimised); 
3. specify a quantity of interest, generally called the figure of merit (FoM),
associated with the proposed experiment. 
We want to extremise the FoM subject to constraints imposed
by the experiment or by our knowledge about the nature of the universe.

Let us assume $e$ denotes the different experimental designs, $M^{i}$ are the different models with
their parameters $\theta^{i}$ and experiment $o$ has already been
performed (this experiment's posterior $P(\theta|o)$ forms
our prior probability function for the new experiment). The FoM (sometimes called the utility function) will depend on the parameters of interest, the
previous experiment (data) and the characteristics of the future
experiment; $U(\theta,e,o)$. From this, we can build
the expected utility $E\left[U\right]$ as
\begin{equation}
E[U|e,o]=\sum_{i}P(M^{i}|o)\int d\hat{\theta}^{i}\; U(\hat{\theta}^{i},e,o)P(\hat{\theta}^{i}|o,M^{i})\:,
\end{equation}
where $\hat{\theta}^{i}$ represent the fiducial parameters for model
$M^{i}$. Our knowledge of the universe is described by the current
posterior distribution $P(\hat{\theta}|o)$. Averaging the utility
over the posterior accounts for the present uncertainty in the parameters
and summing over all the available models would account for the uncertainty
in the underlying true model.
The aim is to select an experiment that extremises the utility function (or its expectation).
One of the common choices for the FoM is some scalar function
of the Fisher matrix, which is the expectation of the inverse covariance
of the parameters in the Gaussian limits (this will be explained in the next
section)\footnote{One can refer to the Dark Energy Task Force (DETF) \citep{DETF} FoM, that use Fisher-matrix techniques to investigate how 
well each model experiment would be able to restrict the dark energy parameters $w_0$, $w_a$, $\Omega_{DE}$ for their purposes.}. Three common FoMs \citep{BayesianBook}, which we will be using as well, are
\begin{itemize}
\item A-optimality $=\log(\textrm{trace}(\mathbf{F}))$;
trace of the Fisher matrix $\mathbf{F}$ (or its $\log$), which is proportional to
sum of the variances. 
\item D-optimality $=\log\left(\left|\mathbf{F}\right|\right)$;
determinant of the Fisher matrix $\mathbf{F}$ (or its $\log$), which measures
the inverse of the square of the parameter volume enclosed by the
posterior. 
\item Entropy (also called the Kullback-Leibler divergence) 
\begin{eqnarray}
\textrm{Entropy} & = & \int d\theta\; P(\theta|\hat{\theta},e,o)\log\frac{P(\theta|\hat{\theta},e,o)}{P(\theta|o)}\nonumber \\
 & = & \frac{1}{2}\left[\log\left|\mathbf{F}\right|-\log|\mathbf{\Pi}|-\textrm{Tr}(\mathbb{I}-\mathbf{\Pi}\mathbf{F}^{-1})\right]\,,
\end{eqnarray}
where $P(\theta|\hat{\theta},e,o)$ is the posterior distribution
with Fisher matrix $\mathbf{F}$ and $P(\theta|o)$ is the prior distribution
with Fisher matrix $\mathbf{\Pi}$. The posterior Fisher matrix is $\mathbf{F}=\mathbf{L}+\mathbf{\Pi}$,
where $\mathbf{L}$ is the likelihood Fisher matrix, which is the
current sparse survey we have designed. Here, the FoMs are defined so that they need to be maximised for an optimal design. For a detailed comparison between the above FoMs please refer to \citet{BayesianBook} and \citet{PaykJaffeSSS}. Note that these are not the `expected' utility functions --- in our current models of the universe, we do not expect a significant difference between the parameters of the same model. 
\end{itemize}

The Fisher matrix \citep{KendallStuart,tegmark1997} has been largely used
for optimisation and forecasting. The Fisher matrix is
defined as the ensemble average of the \textit{curvature} of the likelihood
function $\mathcal{L}$ (i.e., it is the average of the curvature
over many realisations of signal and noise);
\begin{eqnarray}
F_{ij} = \left\langle \mathcal{F}_{ij}\right\rangle 
=\left\langle -\frac{\partial^{2}\ln\mathcal{L}}{\partial\theta_{i}\partial\theta_{j}}\right\rangle \label{eq:General_FM} 
=\frac{1}{2}\textrm{Tr}[C_{,i}C^{-1}C_{,j}C^{-1}]\:,
\end{eqnarray}
where the third equality is appropriate for a Gaussian
distribution with correlation matrix $C$ determined by the parameters
$\theta_{i}$. The
inverse of the Fisher matrix is an approximation of the covariance
matrix of the parameters, by analogy with a Gaussian distribution
in the $\theta_{i}$, for which this would be exact. The Cramer-Rao
inequality%
\footnote{It should be noted that the Cramer-Rao inequality is a statement about
the so-called ``Frequentist'' confidence intervals and is not strictly
applicable to ``Bayesian'' errors. %
} states that the smallest frequentist one-sigma error measured by any unbiased estimator is $1/\sqrt{F_{ii}}$ (non-marginalised) and $\sqrt{(F^{-1})_{ii}}$ (marginalised)\footnote{Integration of the joint probability over other parameters.%
}. The derivatives in Equation \ref{eq:General_FM}
generally depend on where in the parameter space they are calculated
and hence the Fisher matrix is function of the fiducial
parameters. We further note, as in all uses of the Fisher matrix, that any results
thus obtained must be taken with the caveat that these relations only
map onto realistic error bars in the case of a Gaussian distribution,
usually most appropriate in the limit of high signal-to-noise ratio, so that the conditions of the central
limit theorem obtain. In case of no extremely degenerate
parameter directions, we expect that our results will be
indicative of a full analysis \citep{T-BayesExp}.

Following \citep{tegmark1997}, the data in pixel $i$ is defined as 
$
\Delta_{i}\equiv\int d^{3}x \; \psi_{i}(\underbar{x})\left[{n(\underbar{x})-\bar{n}}\right]/{\bar{n}}\,,
$
where $n(\underline{x})$ is the galaxy density at position $\underline{x}$
and $\bar{n}$ is the expected number of galaxies at that position.
The weighting function, $\psi_{i}(\underline{x})$, which determines
the pixelisation and the shape of the survey, is defined as a set of Fourier pixels
\begin{eqnarray}
\psi_{i}(\underline{x})=S(\underline{x}) \; \frac{e^{\iota\underline{k}_{i}.\underline{x}}}{V}\times\begin{cases}
1 & \,\underline{x}\,\,\textnormal{ inside\,\,\ survey\,\,\ volume}\\
0 & \,\textnormal{otherwise}
\end{cases}\,,\label{eq:weighting_fn}
\end{eqnarray}
where $V$ is the {\it total} volume of the survey and $S(\underline{x})$ is the mask (i.e., design of the survey that, of example, defines the distribution of the observed patches). We design the sparsely sampled area of the sky
as a distribution $n_{p}\times n_{p}$ square patches of size $M\times M$
--- see Figure \ref{fig:Design}. Therefore, the structure of the mask $S$ on the sky is defined as a top-hat in both $x$ and $y$ directions and as a step function in the $z$ direction
\begin{equation}
S(\underline{x}) = \Theta(z) \times \sum_{n,m}\Pi(x-x_{n},y-y_m) \, ,
\end{equation}
where $x_{n}$ and $y_{m}$ mark the centres of the patches in our
coordinate system and the functions are defined as 
\begin{eqnarray}
\Pi(x-x_{n},y-y_m) & = & \begin{cases}
1 & |(x-x_{n},y-y_m)|<M/2 \\
0 & \textnormal{otherwise}
\end{cases}\,,\\
\Theta(z) & = & \begin{cases}
1 & z_\textrm{min} < z < z_\textrm{max} \\
0 & \,\textnormal{otherwise}
\end{cases}\,.
\end{eqnarray}

Dividing the survey volume into sub-volumes $i$, $\Delta_{i}$ is then the fractional over-density in pixel $i$.
Using this pixelisation we can define a covariance matrix as
$
\left\langle \Delta_{i}\Delta_{j}^{*}\right\rangle =C=(C_{S})_{ij}+(C_{N})_{ij}\,,
$
where $C_{S}$ and $C_{N}$ are the signal and noise covariance matrices
respectively and are assumed independent of each other. For generality, we take the complex conjugate of one member of the pair. By equating the number over-density $\left[n(\underbar{x})-\bar{n}\right]/\bar{n}$ to the continuous over-density $\delta(\underbar{x})=\left[\rho(\underbar{x})-\bar{\rho}\right]/\bar{\rho}$, the signal and the noise covariance matrices can be defined as 
\begin{eqnarray}
(C_{S})_{ij} & = & \left\langle \Delta_{i}\Delta_{j}^{*}\right\rangle
 = \int\frac{dk}{(2\pi)^{3}}k^{2}P(k)W_{ij}(k)\, ,\\
 (C_{N})_{ij} & = & \left\langle N_{i}N_{j}^{*}\right\rangle
=\int\frac{dk}{(2\pi)^{3}}k^{2}\frac{1}{\overline{n}}W_{ij}(k)\ ,
\label{eq:C_S_ij1}
\end{eqnarray}
where $\tilde{\psi}_{i}(\underline{k})$ is the Fourier transform
of $\psi_{i}(\underline{x})$ and the window function $W_{ij}(k)$
is defined as the angular average of the square of the Fourier transform
of the weighting function; $W_{ij}(k)=\int d\Omega_{k}\tilde{\psi}_{i}(\underline{k})\tilde{\psi}_{j}^{*}(\underline{k})$. For a full analysis of above equations please refer to \cite{ModernCosmology} or \cite{PaykJaffeSSS}. This prescription gives us a data covariance matrix for a galaxy survey, from which we can obtain a Fisher matrix for the parameters of interest using Equation \ref{eq:General_FM}
above.

\section{Results}

We have chosen a geometrically flat $\Lambda$CDM model with adiabatic
perturbations with a five-parameter model: $\Omega_{m}=0.214$, $\Omega_{b}=0.044$, $\Omega_{\Lambda}=0.742$,
$\tau=0.087$ and $h=0.719$, where $H_{0}=100h\,\textnormal{km}\,\textnormal{s}^{-1}\,\textnormal{Mp}\textnormal{c}^{-1}$.
As explained above, the FoM used are Entropy, A-optimality and D-optimality, where a SDSS-LRG-like survey has been chosen as the prior Fisher matrix $\mathbf{\Pi}$.
\begin{figure}
\includegraphics[scale=.265]{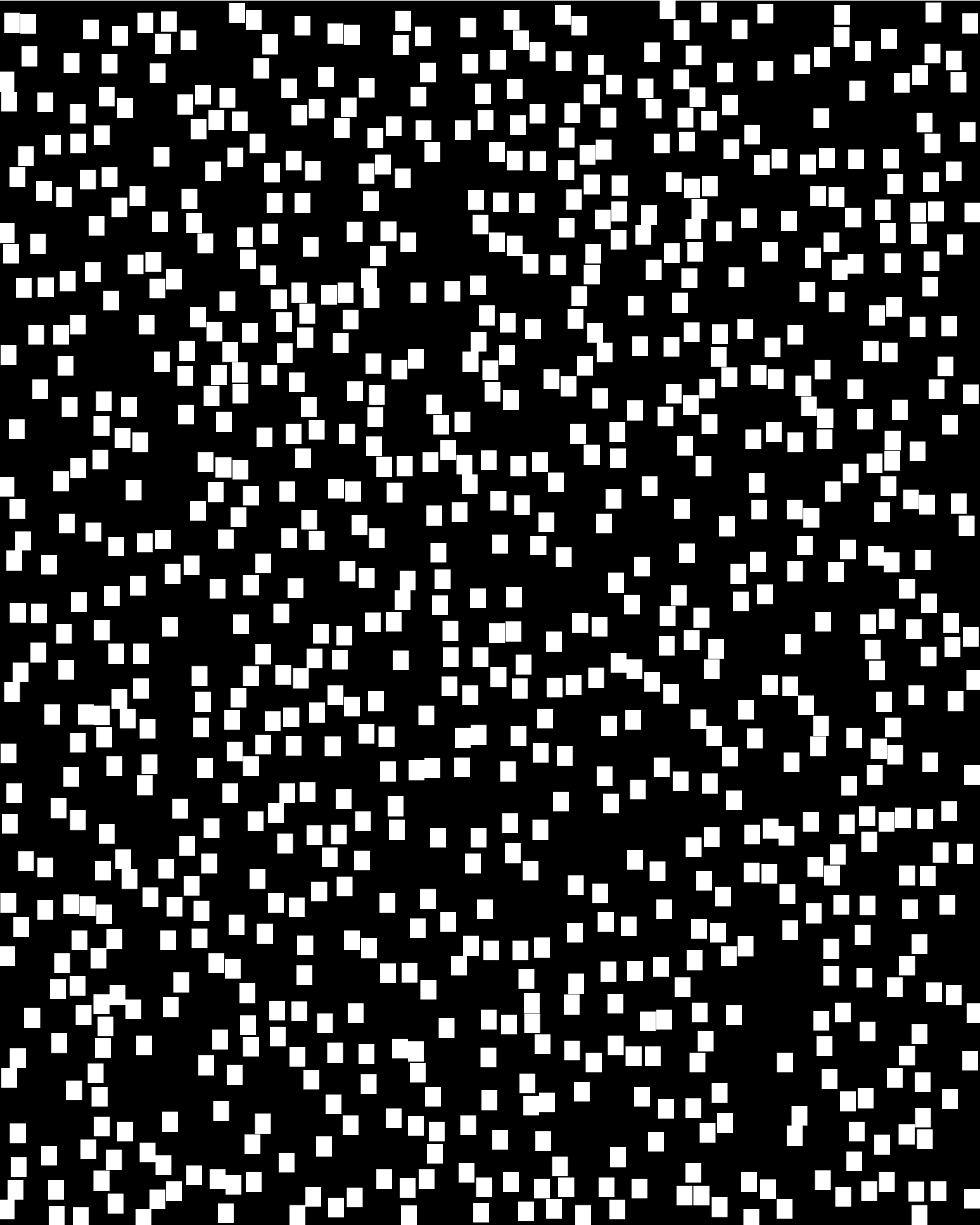}
\includegraphics[scale=.265]{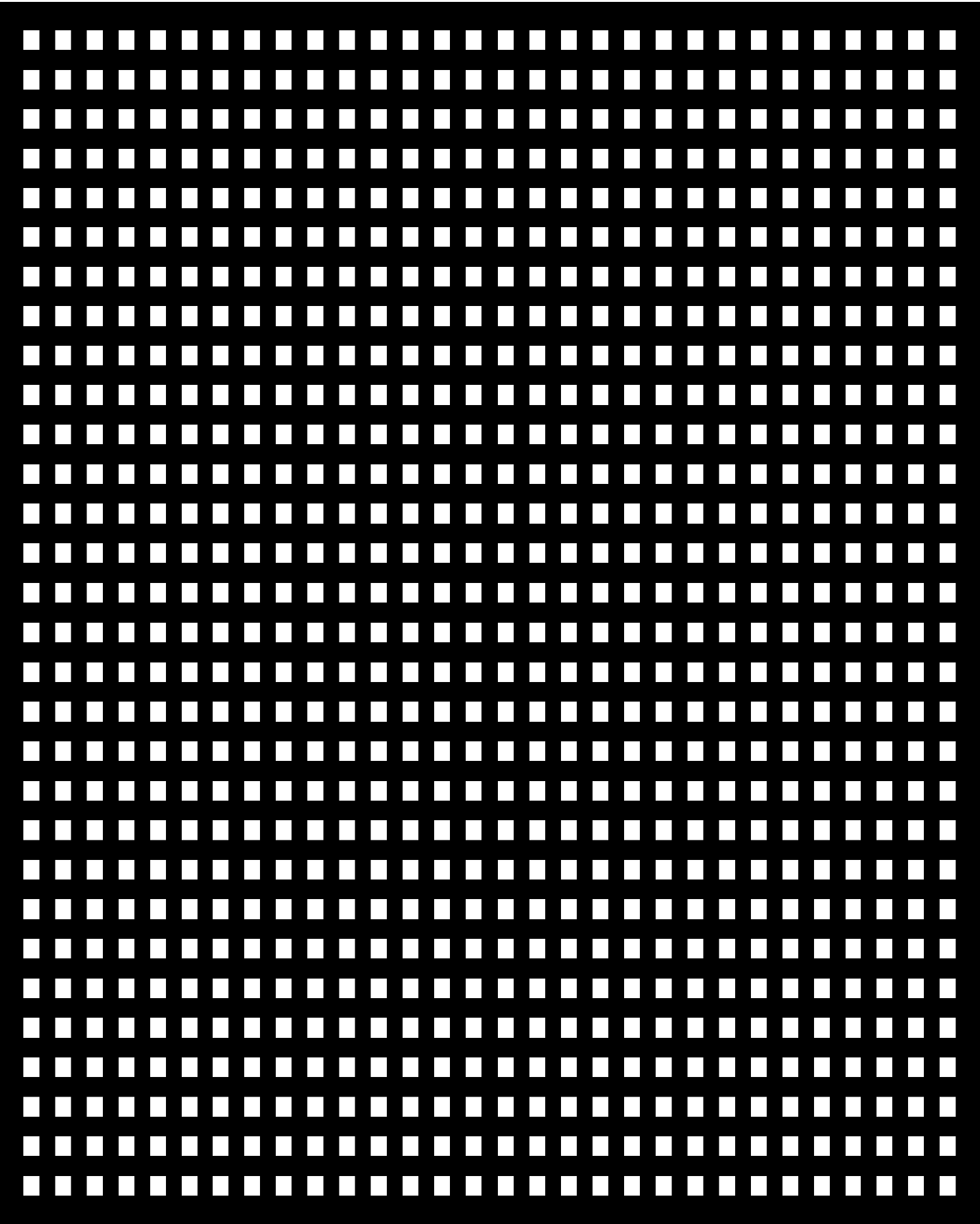}
\caption{Design of the mask for random (left) and regular (right) sampling.  
The patches (we are observing through the white square patches in the Figure), of size $M$, are distributed randomly and regularly on the surface of the sky. The total \textit{observed} area is the sum of the areas of all the patches, $(n_p \times M)^2$, and \textit{total} sampled area is the total area which bounds both the masked and the unmasked areas. Hence the fractional sky coverage is $f_\text{sky}=(n_p \times M)^2/A_{\textrm{tot}}$, which is the same in both designs. 
\label{fig:Design}}
\end{figure}
As in \citet{PaykJaffeSSS}, we use a flat sky approximation to sparsify a survey --- similar to that of the DES survey\footnote{The Dark Energy Survey (DES) \citep{DES} has started taking data in December 2012 and will continue for five years to catalogue 300 million galaxies in the southern
sky over an area of 5000 square degrees and a redshift range of $0.2<z<1.3$.}.
We divide the total sparsely sampled area of the sky
into small square patches and distribute them randomly (left panel of Figure \ref{fig:Design}) and regularly (right panel of Figure \ref{fig:Design}). Note that there are two scales that control the behaviour of the window
function; the size of the patches and the
distance between them. In both designs the size and the number of the patches are kept the same so that the fractional sky coverage, $f_{\text{sky}}$, is the same in both cases. The only difference between the two designs is the distance between the patches due to their different distribution.

\begin{figure*}
\includegraphics[width=\textwidth,height=7.41cm]{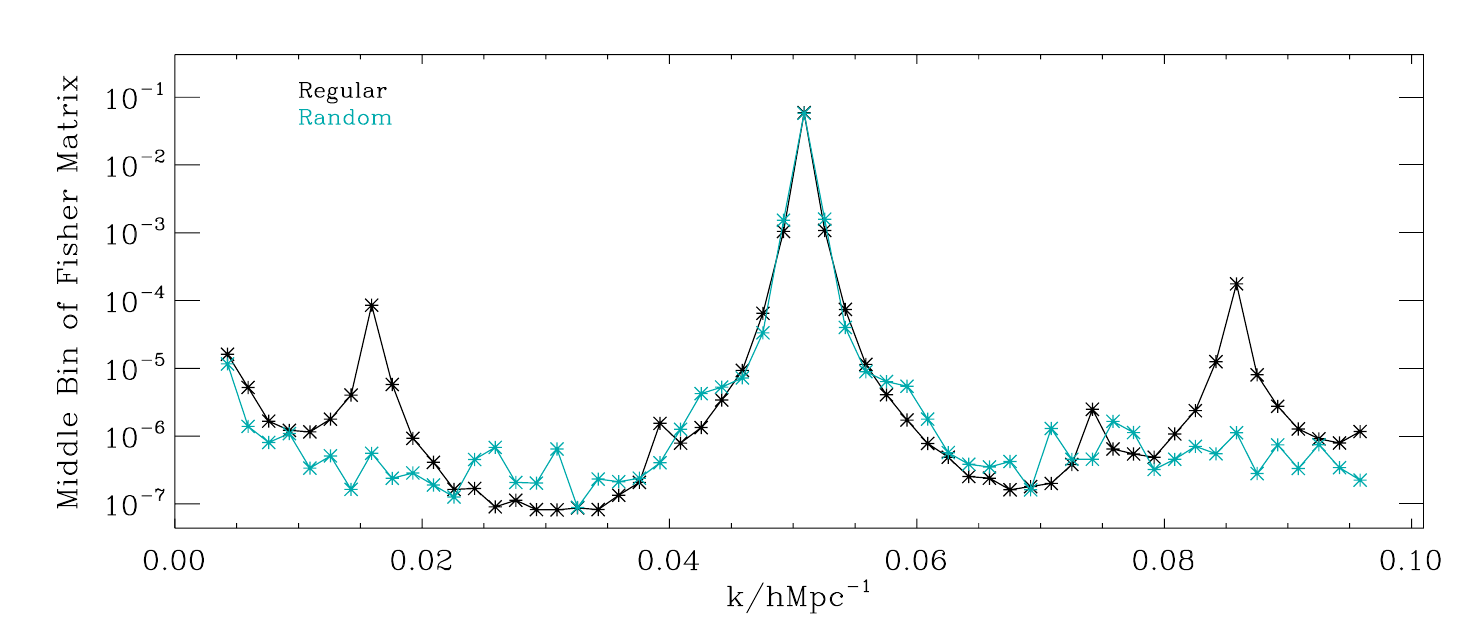}
\caption{The middle row of the power spectrum Fisher matrix for regular (black) and random (blue) sampling. In both cases, the main peak in the middle is the inverse error (remember this is the Fisher matrix) of the middle bin of the power spectrum. Going away from the main peak, each point represents the correlation between that bin and the middle bin. In the case of regular sampling, there are secondary peaks at specific scales,which is due to the fixed position of the patches. On the other hand, for random sampling, there is no preferred scale and correlation is evenly distributed between all scales. The symmetric shoulders on the main peak, at $k\simeq0.037h\,\textrm{Mpc}^{-1}$ and $k\simeq0.06h\,\textrm{Mpc}^{-1}$, in the random case are due to the design of the random mask, which has been obtained by a reflection in the $x$ and $y$ plane for simplicity (see main text).  
Note that the $y$-axis is in log scale.
\label{fig:FM2}}
\end{figure*}

Figure \ref{fig:FM2} shows the middle row of the power spectrum Fisher matrix for regular (black) and random (blue) sampling. In both cases, the main peak in the middle is the expected inverse error of the middle bin of the power spectrum. Going away from the main peak, each point represents the correlation between that bin and the middle one. In the case of regular sampling, apart from the main peak at the centre, there are secondary peaks at other scales indicating an induced correlation at these scales. The position of the secondary peaks is a consequence of the fixed distances between the patches in the mask. The regularity in the mask introduces a periodic pattern in the window function, which in turn induces correlations at that period. Therefore, in this design, certain scales can be less significantly measured than others. This could be a disadvantage if we wish to constrain the behaviour of the power spectrum at a certain scale, such as the BAO scale. On the other hand, in the case of random sampling, the distance between the patches is {\it not} fixed. As there is no preferred scale in the mask, all scales are constrained with almost the same level of significance in this design. This is desirable for constraining the power spectrum at a certain scale, as the power leakage from the main peak is evenly distributed amongst all scales. Also, note that the amplitude and the width of the main peak are controlled by the fractional sky coverage,  $f_\text{sky}$, and the total sparsely sampled volume, $V_\text{tot}$, respectively. As $f_\text{sky}$ and $V_\text{tot}$ are the same in regular and random sampling, the main peaks have same amplitude and width in both cases.

Table \ref{table:FoM} shows the FoM for the galaxy power
spectrum bins on the left and the cosmological parameters on the right. As can be seen, both regular and random designs have very similar values for both the power spectrum bins and the parameters. This shows that, for the same $f_\text{sky}$, the arrangement of the patches does not play an important role in constraining the galaxy spectrum bins or the parameters. Therefore, the constraining power of the survey is not controlled by the distribution of the patches, rather, as investigated in \citet{PaykJaffeSSS}, by the {\it total} extent of the sampled area. However, note that the FoMs we have chosen here do not measure the constraining power of the survey for a particular scale. Rather, they measure the integrated constraining power over all scales of the spectrum. If we wish to measure a particular scale, random sampling would be the preferred approach as it causes an evenly distributed leakage of the power into all the scales.
\begin{table}
\caption{Figure of Merit (FoM) for the galaxy power spectrum and the cosmological parameters for regular and random sampling.  The FoMs are defined so that they need to be maximised for an optimal design.}
\centering  
\begin{tabular}{c c c c c} 
\hline                      
 FoM &\multicolumn{2}{c |}{Power Spectrum} &\multicolumn{2}{c |}{Cosmological Parameters} \\ [0.5ex]
\hline\hline  
 & Regular & Random & Regular & Random \\ [0.5ex] 
\hline                  
Entropy & 39.3 & 39.2 & 3.5 & 3.5  \\ 
D-optimality & -1415.5 & -1415.9 & 11.2 & 11.3  \\
A-optimality & -20.6 & -20.6 & 7.8 & 7.8 \\ [1ex]  
\hline 
\end{tabular}
\label{table:FoM} 
\end{table}

To this end we summarise the main features of Fig.~\ref{fig:FM2}:\\ 
1. The width of the main peak in both designs (and the secondary peaks in the regular case) is controlled by the {\it total} size of the survey. As this is the same in both designs, the width is the same in both cases.\\
2. The position of the secondary peaks in the regular case is controlled by the position of the patches in the mask. Note that apart from the periodicity in the $x$ and $y$ directions, there is also periodicity along the $45^{\circ}$ line. This causes the smaller secondary peaks, for e.g., at $k\simeq0.04$ and $k\simeq0.075h\,\textrm{Mpc}^{-1}$.\\
3. The size of the patches generates an {\it envelope function} over the whole $k$ range. As the size of the patches are so much smaller than the total size of the survey their effect over our $k$ range is negligible. Also, the patches have the same size in both designs, so their effect in the window function is exactly the same.\\
4. As the random mask has been designed as a reflection of a smaller random mask in $x$ and $y$, and is hence not completely randomised over the whole area, some regularities are expected. For example, the symmetric shoulders at $k\simeq0.037h\,\textrm{Mpc}^{-1}$ and $k\simeq0.06h\,\textrm{Mpc}^{-1}$ on the main peak of the random case is due to the patches placed at the edges of the mask. Closer patches also have an effect with a smaller amplitude over a larger range of $k$. This effect is below the remaining small random fluctuations.\\
5. As $f_\text{sky}$ is the same in both designs, the total information gained in both surveys is the same. Note that an average over the positions of the patches in the mask is constant: 
\begin{eqnarray}
\left<S(\underline{x})\right>_{nm} &\sim& \left<\Pi(x-x_{n},y-y_m)\right>_{nm}\,,\nonumber\\
&\sim& \int dx_n\,dy_m\,p(x_n)p(y_m)\, \Pi(x-x_{n},y-y_m)\,,\nonumber\\
&\sim& \frac{1}{4XY}\int_{-X}^{X} \int_{-Y}^{Y} dx_n\, dy_m\, \Pi(x-x_{n},y-y_m)\,,\nonumber\\
&\sim&\textrm{Constant} \,,
\end{eqnarray}
where $2X$ and $2Y$ are the total extent of the survey in the $x$ and $y$ directions and $p(x_n)$ and $p(y_m)$ are the probability distribution of the patches in the $x$ and $y$ directions. Therefore, in terms of the FoM, both designs have the same constraining power for the galaxy power spectrum and the cosmological parameters. 

\section{Conclusion}

For future surveys, one would like to know the optimal investment
of time and money. In the current era, where statistical errors have been greatly reduced
and compete with systematic errors, observing
a greater number of galaxies (to overcome the Poisson noise) may not necessarily improve our results.
One desires more strategic ways to make observations and
take control of systematics. 
This inspired a new approach in making observations \citep[see][]{PaykJaffeSSS}, in which the sampled area was covered sparsely as opposed to contiguously. In this
case one gathers a larger density of states in Fourier space,
but at the expense of an increased correlation between different scales
--- aliasing. This would smooth out features on certain scales and decrease
their statistical significance. In that work, the area of the sky was divided into small square patches, regularly distributed across the total area. It was shown that the loss of the constraining power of the survey induced by the sparse sampling is negligible.

More interestingly, it was shown that for the same amount of observing time, one could sparsely sample a larger total area of sky, which improves the constraining power of the survey. One therefore gains a great deal by spending the same amount of time on a larger but sparsely sampled area. Hence the sparse sampling could be a promising substitute for the contiguous observations and the way forward for designing future surveys. 
However, one constraint in this previous design was the fixed and determined positions of the observed patches. 
The regular design of the mask introduces a periodic pattern in the window function, which induces periodic correlations at specific scales corresponding to the distances between the patches. This is can be a problem if we are interested in a specific scale in in the power spectrum. 

In this work, we have compared the random sampling of sky to regular sampling. In the random design, as there is no preferred scale in the mask, we find that all scales are constrained with almost the same level of significance. 
 
Moreover, in terms of constraining the power spectrum over all scales and constraining the cosmological parameters, there is no difference between regular or random sampling. Therefore, the arrangement of the patches does not control the constraining power of the surveys for the galaxy spectrum or parameter measurements. This means we can relax the regular-grid condition in the sparse mask and any pattern that is practically more suitable can be applied. This is good news because, in practice, it is hard to have a regular mask as there always are regions in the sky one would like to avoid, such as the plane of the Milky Way.

\section{Acknowledgements}

The authors would like to thank A. Woiselle and F. Lanusse for the kind discussions. This work was supported by the European Research Council grant SparseAstro (ERC-228261).

\bibliographystyle{mn2e}
\bibliography{biblio}

\begin{thebibliography}{}

\bibitem[\protect\citeauthoryear{{Albrecht}, {Bernstein}, {Cahn}, {Freedman},
  {Hewitt}, {Hu}, {Huth}, {Kamionkowski}, {Kolb}, {Knox}, {Mather}, {Staggs} \&
  {Suntzeff}}{{Albrecht} et~al.}{2006}]{DETF}
{Albrecht} A.,  {Bernstein} G.,  {Cahn} R.,  {Freedman} W.~L.,  {Hewitt} J.,
  {Hu} W.,  {Huth} J.,  {Kamionkowski} M.,  {Kolb} E.~W.,  {Knox} L.,  {Mather}
  J.~C.,  {Staggs} S.,    {Suntzeff} N.~B.,  2006, ArXiv Astrophysics e-prints

\bibitem[\protect\citeauthoryear{{Dodelson}}{{Dodelson}}{2003}]{ModernCosmology}
{Dodelson} S.,  2003, {Modern cosmology}

\bibitem[\protect\citeauthoryear{{Hobson}, {Jaffe}, {Liddle}, {Mukherjee} \&
  {Parkinson}}{{Hobson} et~al.}{2009}]{BayesianBook}
{Hobson} M.~P.,  {Jaffe} A.~H.,  {Liddle} A.~R.,  {Mukherjee} P.,
  {Parkinson} D.,  2009, {Bayesian Methods in Cosmology}

\bibitem[\protect\citeauthoryear{{Kendall} \& {Stuart}}{{Kendall} \&
  {Stuart}}{1977}]{KendallStuart}
{Kendall} M.,  {Stuart} A.,  1977, {The advanced theory of statistics. Vol.1:
  Distribution theory}

\bibitem[\protect\citeauthoryear{{Liddle}, {Mukherjee} \& {Parkinson}}{{Liddle}
  et~al.}{2006}]{Liddleetal06}
{Liddle} A.,  {Mukherjee} P.,    {Parkinson} D.,  2006, Astronomy and
  Geophysics, 47, 040000

\bibitem[\protect\citeauthoryear{{Paykari} \& {Jaffe}}{{Paykari} \&
  {Jaffe}}{2012}]{PaykJaffeSSS}
{Paykari} P.,  {Jaffe} A.~H.,  2012, ArXiv e-prints

\bibitem[\protect\citeauthoryear{{Peebles}}{{Peebles}}{1973}]{Peebles1973}
{Peebles} P.,  1973, APJ, 185, 413

\bibitem[\protect\citeauthoryear{{Tegmark}}{{Tegmark}}{1997}]{tegmark1997}
{Tegmark} M.,  1997, Physical Review Letters, 79, 3806

\bibitem[\protect\citeauthoryear{{The Dark Energy Survey Collaboration}}{{The
  Dark Energy Survey Collaboration}}{2005}]{DES}
{The Dark Energy Survey Collaboration} 2005, ArXiv Astrophysics e-prints

\bibitem[\protect\citeauthoryear{{Trotta}}{{Trotta}}{2007a}]{Trotta07}
{Trotta} R.,  2007a, MNRAS, 378, 72

\bibitem[\protect\citeauthoryear{{Trotta}}{{Trotta}}{2007b}]{Trotta07-BayesFactor}
{Trotta} R.,  2007b, MNRAS, 378, 819

\bibitem[\protect\citeauthoryear{{Trotta}}{{Trotta}}{2007c}]{T-BayesExp}
{Trotta} R.,  2007c, MNRAS, 378, 819

\end{thebibliography}
\label{lastpage}
\end{document}